\documentclass[apj]{emulateapj}

\usepackage{color}
\usepackage{CJKutf8}
\usepackage{subfigure}
\usepackage[colorlinks=true,citecolor=blue,linkcolor=blue]{hyperref}

\expandafter\def\csname 0532 \endcsname{B0532$-$67.5}
\newcommand {\snr}[1]{\csname #1 \endcsname}
\newcommand {\msun}{$M_\odot$}

\begin{document}

\title{The Shellless SNR B0532$-$67.5 in the Large Magellanic Cloud}

\author{Chuan-Jui Li \begin{CJK}{UTF8}{bsmi}(李傳睿)\end{CJK}\altaffilmark{1}, 
You-Hua Chu \begin{CJK}{UTF8}{bsmi}(朱有花)\end{CJK}\altaffilmark{1}, Chen-Yu Chuang\begin{CJK}{UTF8}{bsmi}(莊鎮宇)\end{CJK}\altaffilmark{1},\\
Guan-Hong Li \begin{CJK}{UTF8}{bsmi}(李冠宏)\end{CJK}\altaffilmark{1,2}
}
 
\affil{$^1$ Institute of Astronomy and Astrophysics, Academia Sinica, No.\ 1, Sec. 4, Roosevelt Rd., Taipei 10617, Taiwan\ 
\\ cjli@asiaa.sinica.edu.tw, yhchu@asiaa.sinica.edu.tw\\
$^2$ Graduate Institute of Astrophysics, National Taiwan University, Taipei 10617, Taiwan.
}

%

\begin{abstract}
The SNR B0532$-$67.5 in the Large Magellanic Cloud (LMC) was first diagnosed 
by its nonthermal radio emission and its SNR nature was confirmed by diffuse 
X-ray emission; however, no optical SNR shell is detected. The OB association
LH75, or NGC\,2011, is projected within the boundary of this SNR. 
We have analyzed the massive star population in and around SNR B0532$-$67.5:
using optical photometric data to construct color-magnitude diagrams (CMDs), 
using stellar evolutionary tracks to estimate stellar masses, and using 
isochrones to assess the stellar ages. From these analyses, we find a 20--25 Myr
population in LH75 and a younger population less than 10 Myr old to the southwest of
LH75.  The center of SNR B0532$-$67.5 is located closer to the core of LH75 than
the massive stars to its southwest.
We conclude that the SN progenitor was probably a member of LH75 with an initial
mass $\sim$15 $M_\odot$.
The SN exploded in an \ion{H}{1} cavity excavated by the 
energy feedback of LH75. The low density of the ambient medium prohibits the 
formation of a visible nebular shell.
Despite the low density in the ambient medium, physical properties of the
hot gas within the SNR interior do not differ from SNRs with a visible shell 
by more than a factor of 2--3. The large-scale \ion{H}{1} map shows that 
SNR B0532$-$67.5 is projected in a cavity that appears to be connected with
the much larger cavity of the supergiant shell LMC-4.
\end{abstract}

\subjectheadings{ISM: supernova remnants --- ISM: individual objects (SNR B0532$-$67.5) --- Magellanic Clouds}

\section{Introduction}  \label{sec:introduction}

Supernova remnants (SNRs) are commonly diagnosed by their diffuse X-ray
emission, nebular shells with elevated [\ion{S}{2}]/H$\alpha$ line ratios, 
and nonthermal radio spectral indices. 
These signatures are associated with SNR shocks going though the ambient
medium, and their detection depends on the density of the medium and the 
sensitivity of observing facilities.
For example, the XMM-Newton X-ray Observatory is able to detect the diffuse
X-rays from an SNR whose ambient density is $>10^{-3}$ H-atom cm$^{-3}$,
while the commonly available CCD cameras with a narrow H$\alpha$ filter 
($\Delta\lambda \sim 10-20$ \AA) can detect an SNR shell only if the 
ambient density is greater than $\sim$1 H-atom cm$^{-3}$ \citep{Ou2018}. 
It is thus not surprising that some SNRs are diagnosed in X-rays but 
could not be confirmed in optical wavelengths.
An outstanding example is the SNR B0532$-$67.5 in the Large Magellanic Cloud
(LMC), the subject of this paper.
We have chosen to study the LMC SNRs because their resolved underlying stellar 
population can be used to infer properties of suprenova (SN) progenitors, especially 
for the core-collapse SNe.

SNR B0532$-$67.5 was first reported by \citet{Mathewson1985}, who noted the absence 
of an associated optical shell structure, despite the presence of nonthermal 
radio emission and X-ray source.
The diffuse X-ray emission from B0532$-$67.5 has been confirmed by ROSAT, Suzaku, 
and XMM-Newton observations and their spectral analyses show an interstellar 
abundance and an X-ray luminosity of a few $\times$10$^{35}$ erg s$^{-1}$ 
\citep{Maggi2016, Takeuchi2016, Ramirez-Ballinas2019}.
Far Ultraviolet Spectroscopic Explorer (FUSE) observations of B0532$-$67.5 did not detect
the O VI $\lambda\lambda$ 1032, 1037 line emission \citep{Blair2006}.
Spitzer Space Telescope observations did not detect any near- or mid-IR dust 
emission from this SNR \citep{Seok2013}, and the Herschel Space Observatory did 
not detect mid- to far-IR dust emission, either \citep{Lakievic2015}.
It has been noted that SNR B0532$-$67.5 is likely associated with the OB 
association LH75, also known as NGC\,2011 \citep{Lucke1970, Chu1997}.

To understand the origin and evolution of SNR B0532$-$67.5, we have examined the 
stellar content of LH75 and explored its interstellar environment.  This paper
reports our investigation. 
Section 2 describes the multiwavelength observations used in this study, 
Section 3 examines the stars in LH75 to assess the SN progenitor's 
mass, Section 4 analyzes the interstellar environment 
and its effects on the SNR's physical properties, and Section 5 summarizes 
this work and its conclusions.


\begin{figure*}[ht]  
\begin{center}
\epsscale{1.} 
\plotone{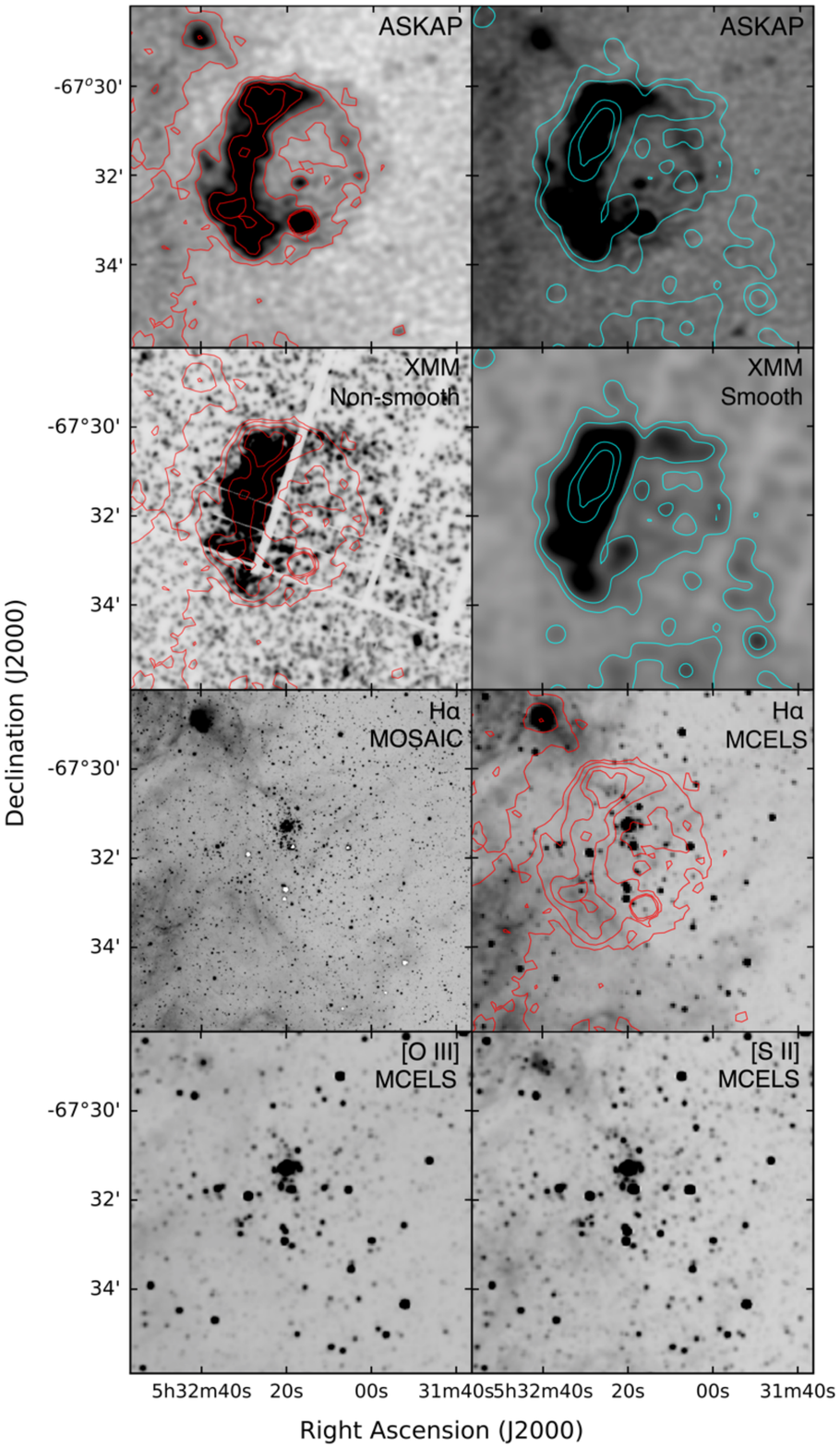} 
\caption{Multiwavelength images of SNR B0532$-$67.5.
The origin of each image is marked in the upper right corner.
The top row presents ASKAP 888 MHz continuum images, and the second
row XMM-Newton X-ray images in the full energy band (0.1--15 keV).  
The bottom four images are CTIO 4m MOSAIC
H$\alpha$, MCELS H$\alpha$, [\ion{O}{3}], and [\ion{N}{2}] images.
To facilitate inter-comparisons, the contours of the ASKAP 888 MHz 
radio continuum image (left panel of the top row) are over-plotted 
on the XMM non-smoothed X-ray image and the MCELS H$\alpha$ image, and the contours of the XMM smoothed 
X-ray image are over-plotted on the ASKAP 888 MHz image (right panel 
of the top row).}
\label{figure:snr0532_line_images}
\end{center}
\end{figure*}

\section{Observations} \label{sec:observations}  

\subsection{Available Photometric Catalogs of Stars}

The $UBVI$ photometric data of LH75 are from the 
Magellanic Cloud Photometric Survey (MCPS), which 
was made with the Great Circle camera on the 
Swope 1\,m Telescope at Las Campanas \citep{Zaritsky2004}. 
These photometric data are used to construct 
color-magnituce diagrams (CMDs) in order to compare 
locations of stars with stellar evolutionary tracks to
assess the massive star content of LH75.  

To complement the MCPS photometry,
We have also used the near-infrared $JHK_s$ photometric 
data from the InfraRed Survey Facility (IRSF) Survey 
\citep{Kato2007} and the Two Micron All Sky Survey 
\citep[2MASS;][]{Skrutskie2006}.
The IRSF Survey of the Magellanic Clouds used the 
SIRIUS camera on the IRSF 1.4\,m 
Telescope at the South African Astronomical Observatory, 
while the 2MASS used its 1.3\,m telescope at Cerro Tololo
Inter-American Observatory (CTIO) for the southern sky observations.

\subsection{Surveys of the ISM} 

To investigate the ionized interstellar gas, we have used images from
the Magellanic Cloud Emission Line Survey \citep[MCELS;][]{Smith1999}.
The MCELS survey used CCD cameras on the Curtis Schmidt 0.61\,cm
Telescope at CTIO and imaged the Magellanic Clouds in the H$\alpha$, [\ion{O}{3}],
[\ion{S}{2}] lines as well as a red continuum and a green continuum.
For better resolution, we have also used an H$\alpha$ image taken with
the MOSAIC camera on the Blanco 4\,m Telescope at CTIO, as shown in 
Figure~\ref{figure:snr0532_line_images},

The \ion{H}{1} interstellar gas in the LMC has been surveyed with the 
Australia Telescope Compact Array (ATCA) in conjunction with the Parkes 
Telescope observations \citep{Kim2003}.  We have used this \ion{H}{1} 
data cube to illustrate the cavities in the neighborhood of SNR B0532$-$67.5
and compare the \ion{H}{1} velocities with those of ionized gas detected
in the echelle observation that is described in the next subsection.

The hot, 10$^6$ K, ionized gas in the SNR B0532$-$67.5  was detected 
in an XMM-Newton X-ray observation (Obs.\ ID 0400320101; 
PI: R. Williams).  This observation has been analyzed and
reported by \citet{Maggi2016} and \citet{Ramirez-Ballinas2019}.  
We will only present an X-ray image to illustrate the extent 
of hot ionized interstellar gas without repeating the X-ray
spectral analysis of the hot gas properties. The exposure times
for the EPIC-pn, MOS1, and MOS2 cameras are 42.6, 46.4, and 46.6 ks,
respectively; however, after filtering out the high-background periods,
the useful exposure times are 7.2, 10.6, and 12.1 ks, respectively
\citep{Ramirez-Ballinas2019}.

\begin{deluxetable*}{cclccccl}
\tablecolumns{7}
\tablewidth{0pc}
\tablecaption{Summary of the ASKAP observational parameters}
\tablehead{Project ID & Sideband ID & ~~Obs Date & Integration & Beam Size \& P.A. & ~~Stokes\\
& & & Time (s) & &}
\startdata
AS107$^a$ & 11125 & 2020 Jan 10 & 727 & $34\arcsec\times32\arcsec$, ~\,~$37\arcdeg$ & I,Q,U,V\\
& 11546 & 2020 Jan 24 & 727 & $34\arcsec\times32\arcsec$, ~~~~$0\arcdeg$ & I,Q,U,V\\
& 11570 & 2020 Jan 25 & 727 & $34\arcsec\times32\arcsec$, ~~~~$0\arcdeg$ & I,Q,U,V\\
& 15665 & 2020 Aug 29 & 717 & $34\arcsec\times32\arcsec$, ~\,~$26\arcdeg$ & I,Q,U,V\\
& 25517 & 2021 Apr 01 & 727 & $34\arcsec\times32\arcsec$, $-10\arcdeg$ & I,Q,U,V\\
AS113$^b$ & 10905 & 2019 Dec 19 & 727 & $34\arcsec\times32\arcsec$, ~\,~$22\arcdeg$ & I,Q,U,V\\
AS033$^c$ & ~8532 & 2019 Apr 20 & 45,476 & $14\arcsec\times12\arcsec$, $-84\arcdeg$ & I
\enddata
\tablenotetext{a}{From \citet{Murphy2020}.}
\tablenotetext{b}{From \citet{Hotan2020}.}
\tablenotetext{c}{From \citet{Harvey-Smith2016}.}
\label{table:askapobs}
\end{deluxetable*}

The Australian Square Kilometre Array Pathfinder (ASKAP) has observed the
general regions around SNR B0532$-$67.5 in the 888 MHz continuum with 
various configurations.  We have found seven archival images that 
can be used to analyze the total intensity and polarization properties 
of SNR B0532$-$67.5.  The relevant parameters of these seven ASKAP
observations are summarized in Table \ref{table:askapobs}.
The top panels of Figure \ref{figure:snr0532_line_images} show the 888 MHz 
continuum image from project AS033. 

\subsection{Echelle Observations of the Ionized Gas}

To examine the kinematic properties of ionized gas around the 
SNR B0532$-$67.5, we obtained a long-slit high-dispersion
spectroscopic observation with the echelle spectrograph 
on the Blanco 4\,m Telescope at CTIO on 2001 December 25.
The  79 l mm$^{-1}$ echelle grating was used.  For the long-slit
single-order observing mode, a broad H$\alpha$ filter was used to 
isolate the order and the cross disperser grating was replaced by
a flat mirror.  The observation used a slitwidth of 250 $\mu$m 
(corresponding to 1\farcs64) and an integration time of 1200 s.  
The data were recorded with the long-red camera and a 
SITe2K\_6 2048$\times$2048 CCD.  The resulting echelle image has 
pixel scales of 0.082 \AA\ pixel$^{-1}$ along the dispersion and 
0\farcs26 pixel$^{-1}$ along the slit.
Both the H$\alpha$ line and the [\ion{N}{2}] $\lambda\lambda$6548, 
6583 doublet are detected.

\section{Assessing SN Progenitor's Mass} \label{sec:LH75}  

The SN progenitor of SNR B0532$-$67.5 may be a member of the
OB association LH75; thus, we analyze the stellar content of
LH75 and the surrounding field stars, in order to determine the
origin of the progenitor and its possible stellar mass.
The OB association LH75, or NGC\,2011, has been studied as
a cluster using either integrated cluster photometry or 
resolved stellar photometry.  
Its age determined from integrated photometry ranges from $\sim$5 Myr 
\citep{Hunter2003} to 14 Myr \citep{Popescu2012}, while its 
age determined from CMDs of resolved stars is higher, 
20 Myr \citep{Kumar2008} or 25 Myr \citep{Glatt2010}.

\begin{figure*}[h]  
\epsscale{1.} 
\plotone{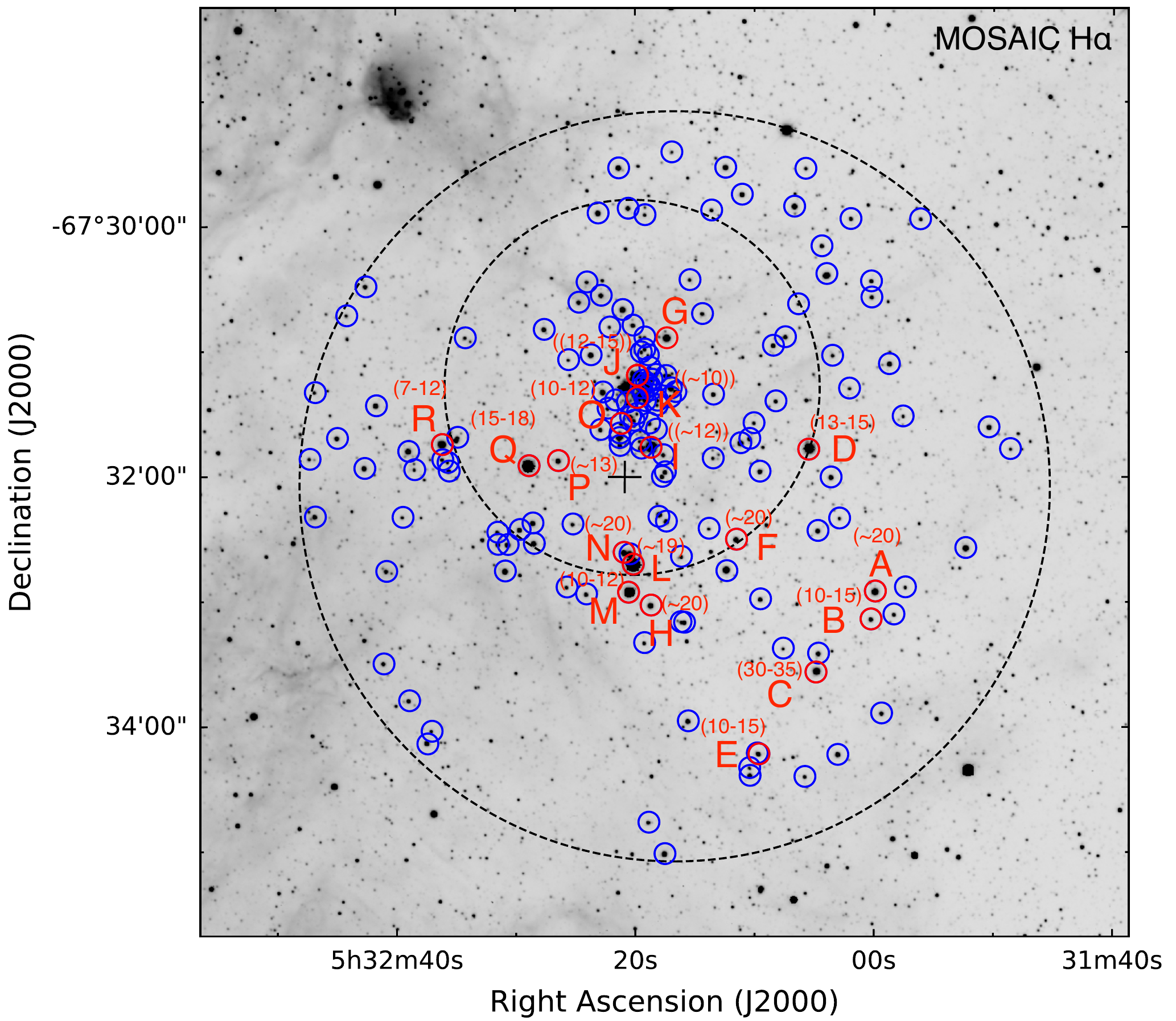} 
\caption{H$\alpha$ image of SNR B0532$-$67.5 showing the star field. A small dashed circle 
with a radius of 1\farcm5 (or 22.5 pc) is drawn around the OB association LH75.  A large 
dashed circle with a radius of 3$'$ (or 45 pc) is drawn around the SNR and encompasses 
the star field analyzed. The center of the SNR is marked by "+". 
Within this field stars with $V < 17$ mag are marked with blue 
circles and used to construct the CMDs.  Candidates for the most massive stars are marked 
with red circles and labeled from A to R. The numbers in parentheses above the labels are 
their estimated masses in units of $M_\odot$. Stars that are multiple or confused have larger 
uncertainties in their masses, and they are marked with double parentheses.} 
\label{figure:star_field}
\end{figure*}

We have examined the MCPS $UBVI$ photometric data, as well as 
the IRSF and 2MASS $JHK_s$ photometric data, of resolved stars in LH75; 
however, we treat it as an OB association and place emphasis on 
only stars that are massive enough to end their lives in SNe.
We made $V$ versus ($B-V$) CMD without extinction correction 
for all stars with $V < 17$ within 3$'$ (or 45 pc) radius from the center 
of the SNR.  Figure \ref{figure:star_field} shows the star field that we
analyzed.  It is clear that the field encompassed the entire OB association 
LH75, whose dimension is 3$'$ (45 pc diameter) according to \citet{Lucke1970}; 
furthermore, from comparisons with Figure~\ref{figure:snr0532_line_images} it 
can be seen that this star field generously encompasses the entire SNR 
B0532$-$67.5. 


\begin{figure*}[h]  
\epsscale{0.9} 
\plottwo{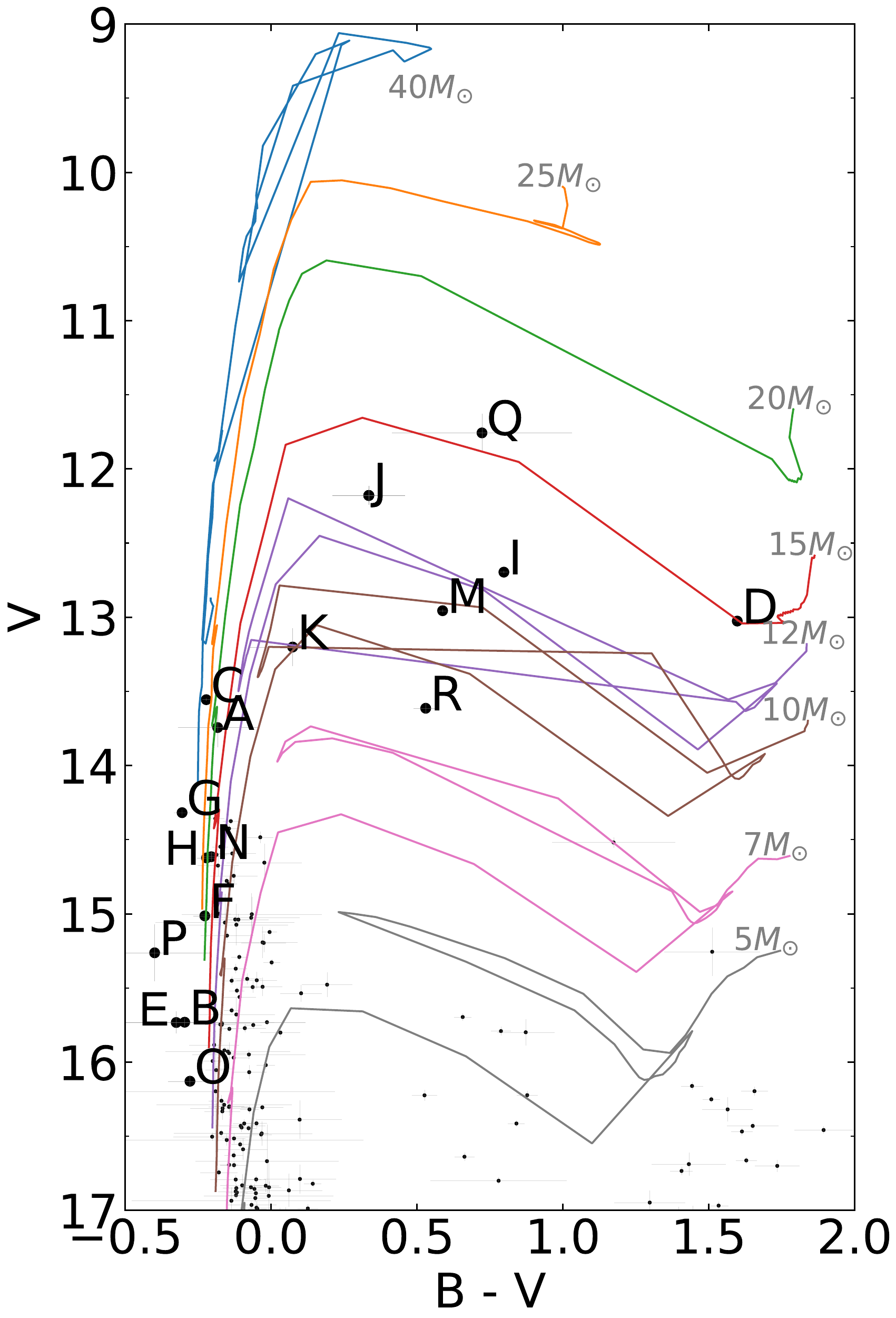}{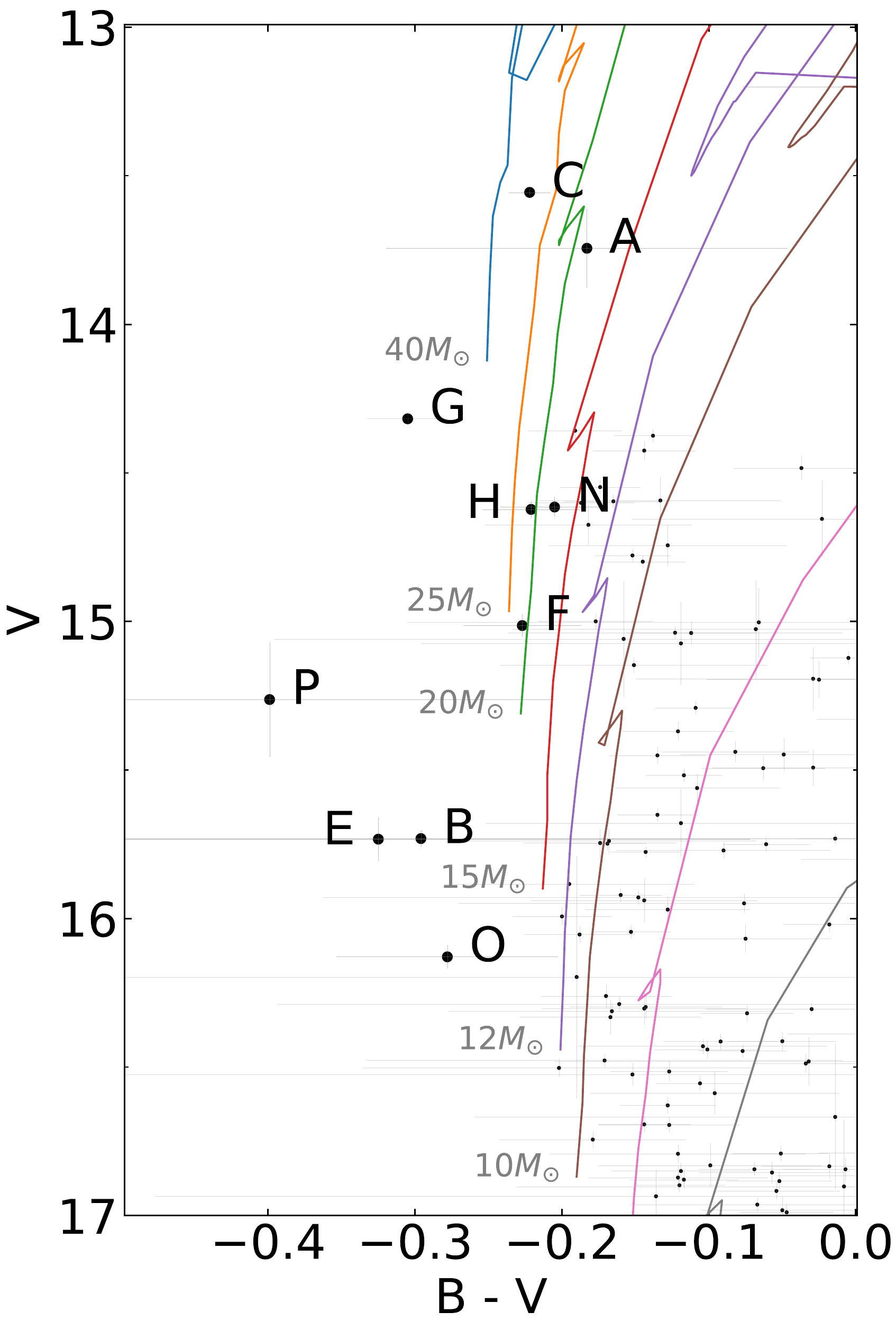}
\caption{Left: $V$ versus $B–V$ CMD of stars with $V <$ 17 mag projected in and around SNR \snr{0532}.
The MCPS photometric data \citep{Zaritsky2004} are used.  The evolutionary tracks of stars 
for $Z = 0.008$ from
\citet{Lejeune2001} are reddened by $E(B-V) = 0.08$ and plotted. The most massive star candidates 
are marked from A to R both in the CMDs and in the image in Figure~\ref{figure:star_field}.
Right: close-up $V$ versus $B–V$ CMD near the main-sequence of evolutionary tracks.
}
\label{figure:CMD_snr0532_MCPS}
\end{figure*}



\begin{figure*}[h]  
\epsscale{0.9} 
\plottwo{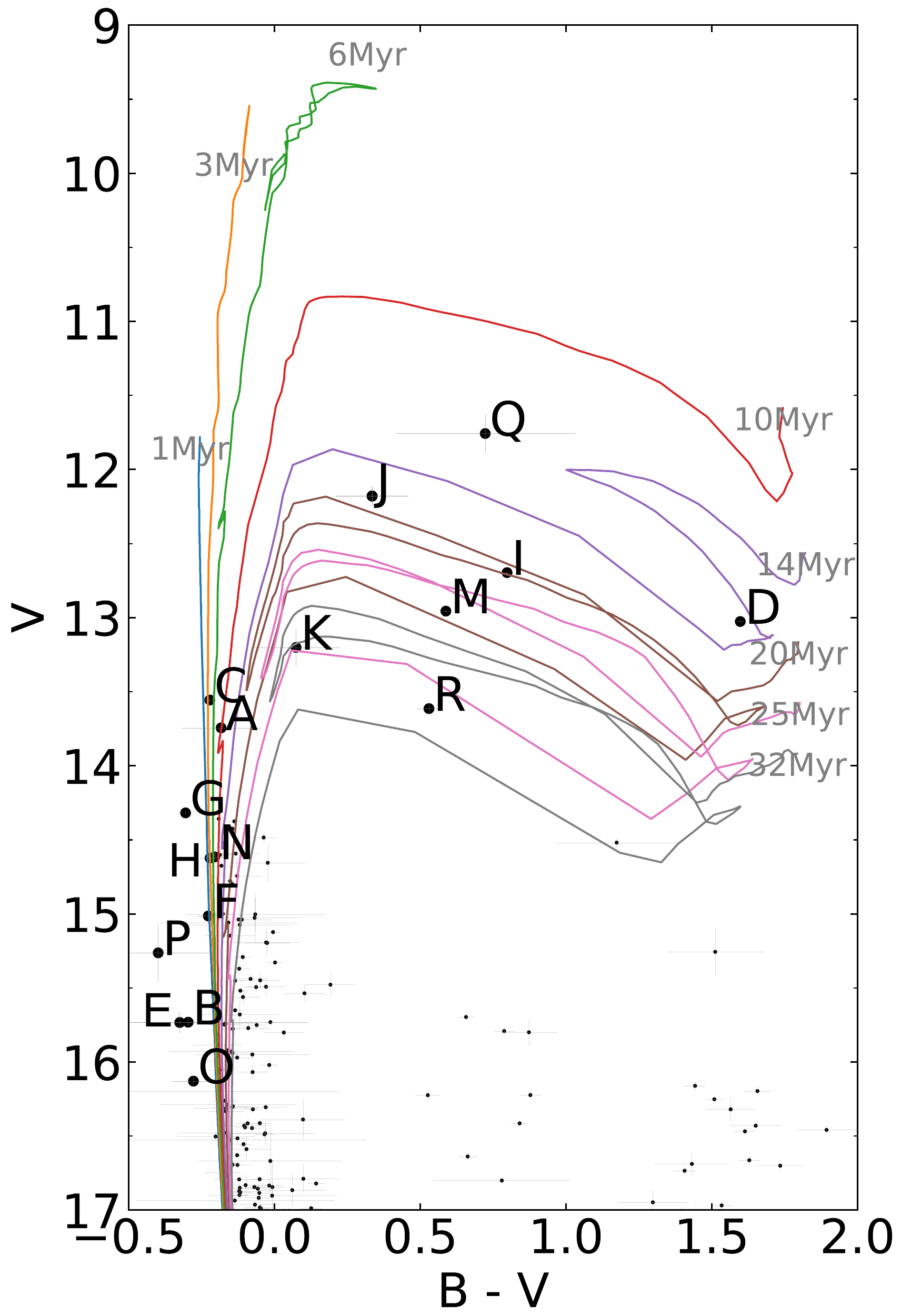}{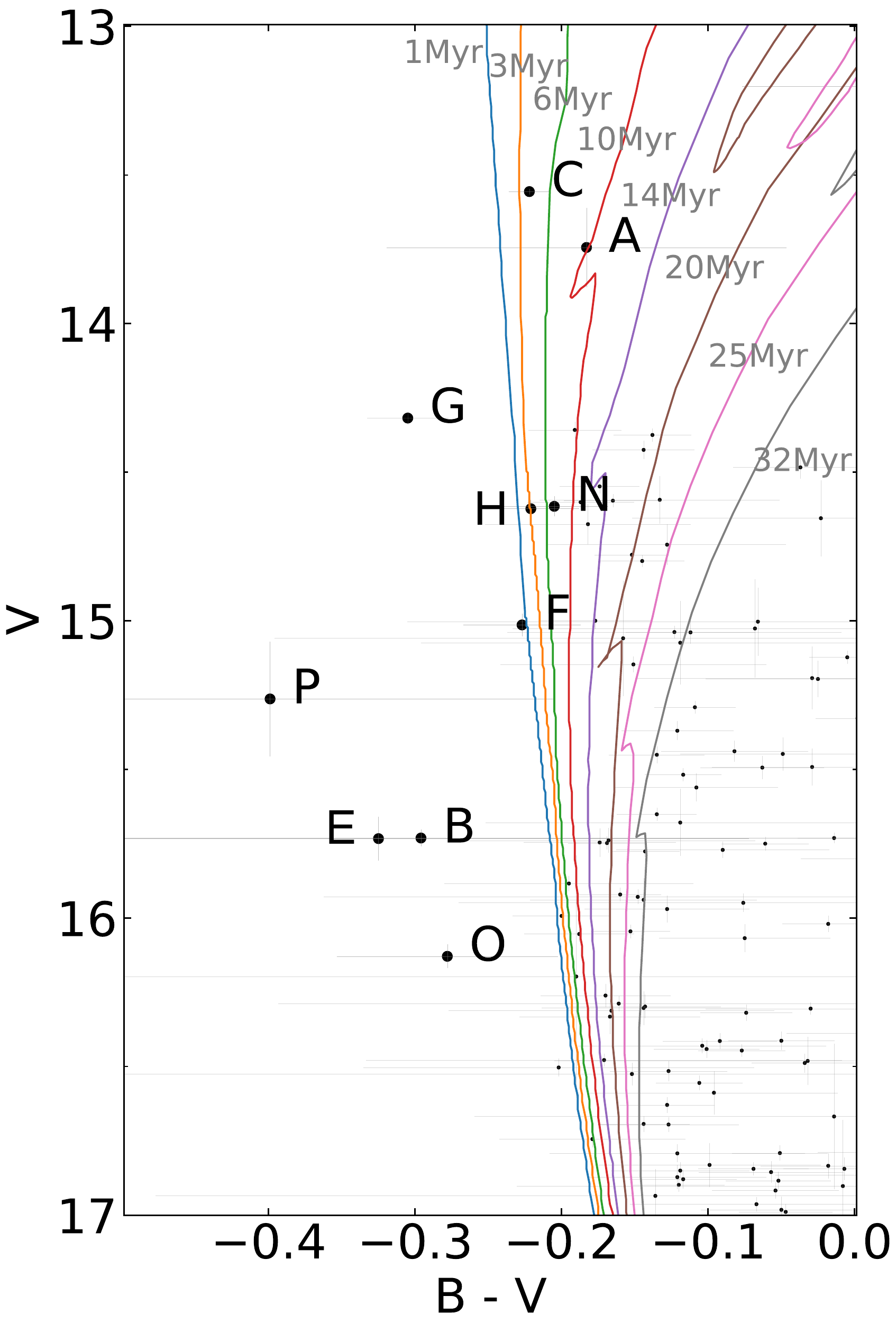} 
\caption{Same as Figure~\ref{figure:CMD_snr0532_MCPS}, except that
isochrones for $Z = 0.008$ from \citet{Lejeune2001}  are 
reddened by $E(B-V) = 0.08$ and plotted.}
\label{figure:CMD_snr0532_MCPS_iso}
\end{figure*}

The $V$ versus ($B-V$) CMD of the above-mentioned star field is presented
in Figure \ref{figure:CMD_snr0532_MCPS}.
The extinction toward LH75 is small, $E(B-V) = 0.08$ \citep{Kumar2008}, but not
negligible; thus, we apply this extinction to the stellar evolutionary tracks 
for $Z = 0.008$ (closest to the LMC metallicity)
from \citet{Lejeune2001} and plot the reddened tracks in Figure \ref{figure:CMD_snr0532_MCPS}.
A closeup of the CMD for massive stars near the main sequence (MS) is presented 
in the right panel of this figure.  
Stars marked from A to R are candidates for the most massive stars in
the field.  They are marked in the H$\alpha$ image in 
Figure \ref{figure:star_field} as well as the CMDs in 
Figure \ref{figure:CMD_snr0532_MCPS}.
These 18 stars' coordinates from MCPS, MCPS $UBVI$, and IRSF or 2MASS $JHK_s$ magnitudes 
are compiled in Table \ref{table:photometry}.
For each star, we compare its $UBV$ and $JHK_s$ colors and magnitudes with those
of standard stars of luminosity classes V and I conveniently tabulated in
\citet{Carroll1996} and \citet{Sparke2000} based on the original data from \citet{SK82}.
If a good match is found, the spectral type is adopted and given in the 11th column
of Table \ref{table:photometry}.
However, if the star has been spectroscopically classified and reported in the literature,
its published spectral type is adopted and noted in the table.

The locations of stars in the $V$ versus ($B-V$) CMD are compared with
the stellar evolutionary tracks to assess the stellar masses.  
These estimated stellar masses are given in the 12th column of 
Table \ref{table:photometry}.
The uncertainty in the mass estimates may be 2--3 $M_\odot$ for stellar masses
below 10 $M_\odot$, but the uncertainty is much larger for high-mass stars and
may be 5 $M_\odot$ or even higher.  
Detailed notes on our estimation of initial mass of each star are  
given in the Appendix.

The most massive star in the star field is C (30--35 $M_\odot$), outside LH75.
The next most massive stars, in the 18--20 $M_\odot$ range, are A, F, H, L, and N.
It is very interesting that these most massive stars are all located to the
southwest of LH75, as seen in Figure~\ref{figure:star_field}.
None of them are near the core of LH75, where stellar masses are almost all $\le 15 M_\odot$.
There is an apparent mass segregation.

Mass segregation in a young star cluster has been predicted from numerical simulations
\citep[e.g.,][]{Dib2010,Haghi2015}; however, massive stars are expected to concentrate at the
center, contrary to what we see in LH75, and observational evidence of mass segregation
in clusters has often been reported to be absent \citep[e.g.,][]{Elson1987,Ascenso2009}.
The reality may be that some clusters show mass segregation and some do not \citep{Schilbach2006}.
In a detailed analysis of the LMC cluster NGC\,1818, \citet{deGrijs2002} show that an 
excess number of bright stars are present around the cluster and suggest massive stars
were ejected from the cluster core.
The mass segregation in NGC\,1818 is similar to what we observe in LH75.

To understand the mass segregation and to assess the relationship among LH75,
massive stars to its southwest, and the SN progenitor, we plot isochrones 
on the CMD in Figure \ref{figure:CMD_snr0532_MCPS_iso} to estimate the ages 
of the stars.  It can be seen that the massive stars (A, C, F, H, L, and N)
to the southwest of LH75 are all less than about 10 Myr old, while stars 
in the core of LH75 (I, J, and K) are 20--25 Myr old.  The MS turnoff
of LH75 is also consistent with an age of 20--25 Myr.
The different ages indicate that the massive stars cannot be ejected from LH75.
In fact, such juxtaposition of two stellar populations with different ages is 
frequently seen in star-forming regions where massive young stellar objects are
detected along the periphery of superbubbles blown by OB associations 
\citep[e.g.,][]{Chen2009,Carlson2012}.
The formation of superbubbles requires energy feedback from both fast winds and SN explosions; 
therefore, there should be several Myr age difference between the central OB association and 
the stars subsequently formed through the superbubble compression of ambient ISM.

The key question to ask is thus whether the SN progenitor of B0532$-$67.5 was a member of
LH75 or one of the massive stars formed through the energy feedback of LH75.
To estimate the site of SN explosion, we first considered the outermost boundary of the X-ray
and radio emission. The shock advances faster in low-density directions and
the large contrast between the bright east rim and the very faint northwest rim indicates 
the explosion site must be closer to the bright east rim. We then considered the 
overall bright radio rim of SNR B0532$-$67.5. Despite the surface brightness variations,
the shape of the bright radio rim is elliptical and regular, and its center can be easily
assessed. We adopt the center of the ellipse as the site of SN explosion and mark it by 
a cross in Figure~\ref{figure:star_field}.  Note that the real SN explosion site may be 
closer to the bright east rim than our adopted center. 
It can be seen that the center of the SNR is closer to the core of LH75 than the group 
of massive stars that are less than 10 Myr old.
Even if the SN site should be closer to the brightest part of the rim, which is east of
LH75 and northeast of our adopted center, the SN site is still closer to LH75 than the group 
of massive stars.
Therefore, we suggest that B0532$-$67.5's SN progenitor was more likely a member of
the OB association LH75 and its mass was close to $\sim$15 $M_\odot$.
In the less likely case that the SN progenitor belonged to the group of massive stars 
to the southwest of LH75, its mass would have been $\gtrsim 20 M_\odot$.

\section{Multiphase Interstellar Environment} \label{sec:ISM} 


\begin{figure*}[th]  
\begin{center}
\epsscale{1.1} 
\plotone{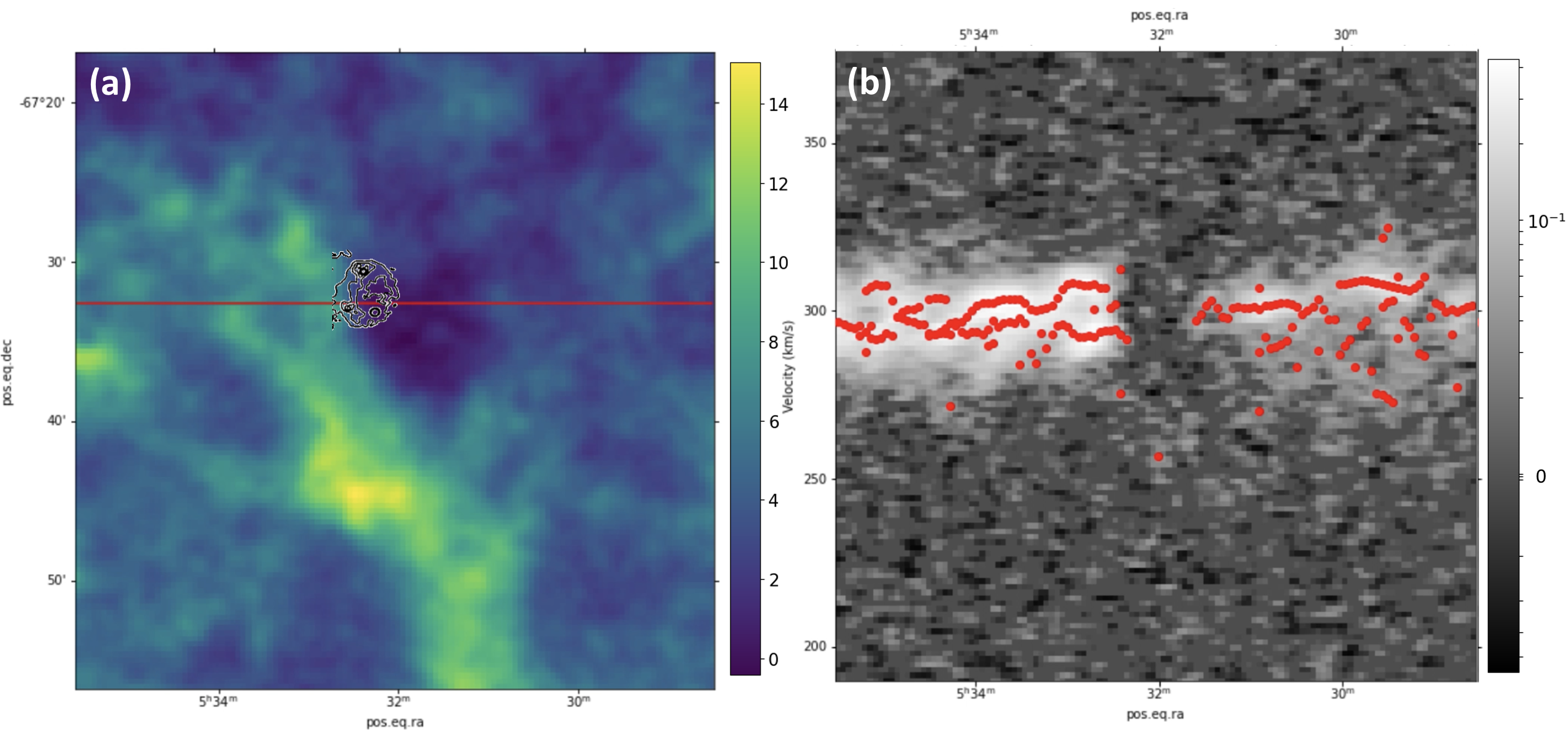} 
\caption{(a) ATCA/Parkes HI moment 0 map (color) integrated over a velocity range of $V_{\rm LSR}=243-348\ $km s$^{-1}$, superposed with contours extracted from the ASKAP 
888 MHz radio continuum image (AS033) at levels of 0.40, 1.55, 2.70, 3.85, and 5.00 mJy beam$^{-1}$. 
The unit of the color bar is Jy beam$^{-1}$ km s$^{-1}$. (b) \ion{H}{1} position-velocity diagram 
along the red line marked in (a). A log scale is used in the image, and the unit of the color bar is
Jy beam$^{-1}$.  The red dots denote the peak positions of each Gaussian 
component in the multi-component Gaussian fitting.}
\label{figure:hispec}
\end{center}
\end{figure*}


\begin{figure*}[th]  
\begin{center}
\epsscale{1.} 
\plotone{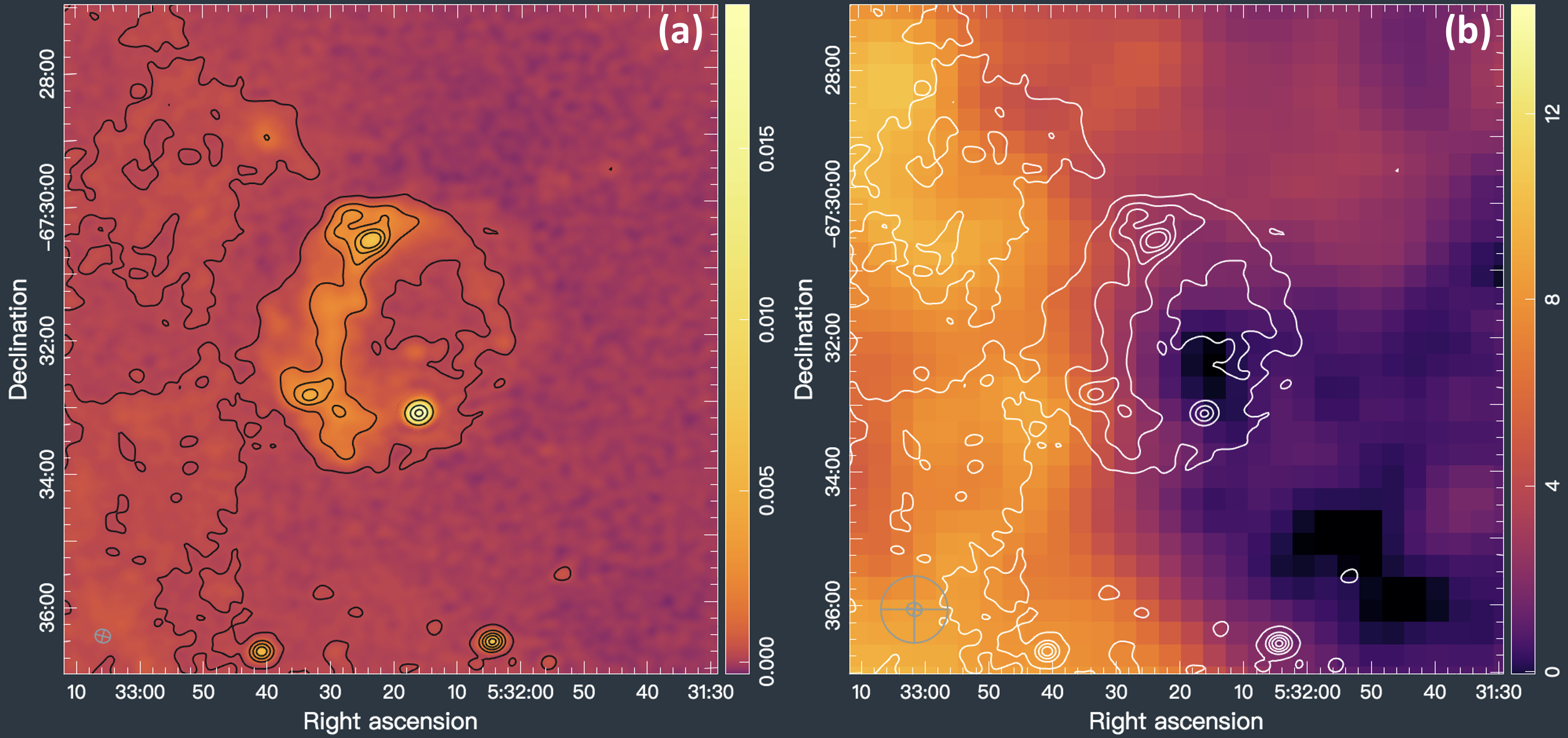} 
\caption{(a) ASKAP radio 888 MHz continuum image of B0532$-$67.5 with contours at levels of 
0.40, 1.55, 2.70, 3.85, and 5.00 mJy beam$^{-1}$.  Three additional contour levels 
of 6.00, 17.0, and 28.0 mJy beam$^{-1}$ are plotted for the bright compact source.
The unit of the color bar is mJy beam$^{-1}$.  (b) A close-up of Figure \ref{figure:hispec}(a) 
centered at SNR B0532$-$67.5. A log scale is used in this image, and the unit of the color bar is 
Jy beam$^{-1}$ km s$^{-1}$.  The radio contours are the same as those in panel (a).}
\label{figure:hiaskap}
\end{center}
\end{figure*}


\begin{figure*}[th]
\begin{tabular}{cc}
\hspace{1.2cm}
\subfigure[]{
\hspace{-2.cm}
\includegraphics[scale=0.73]{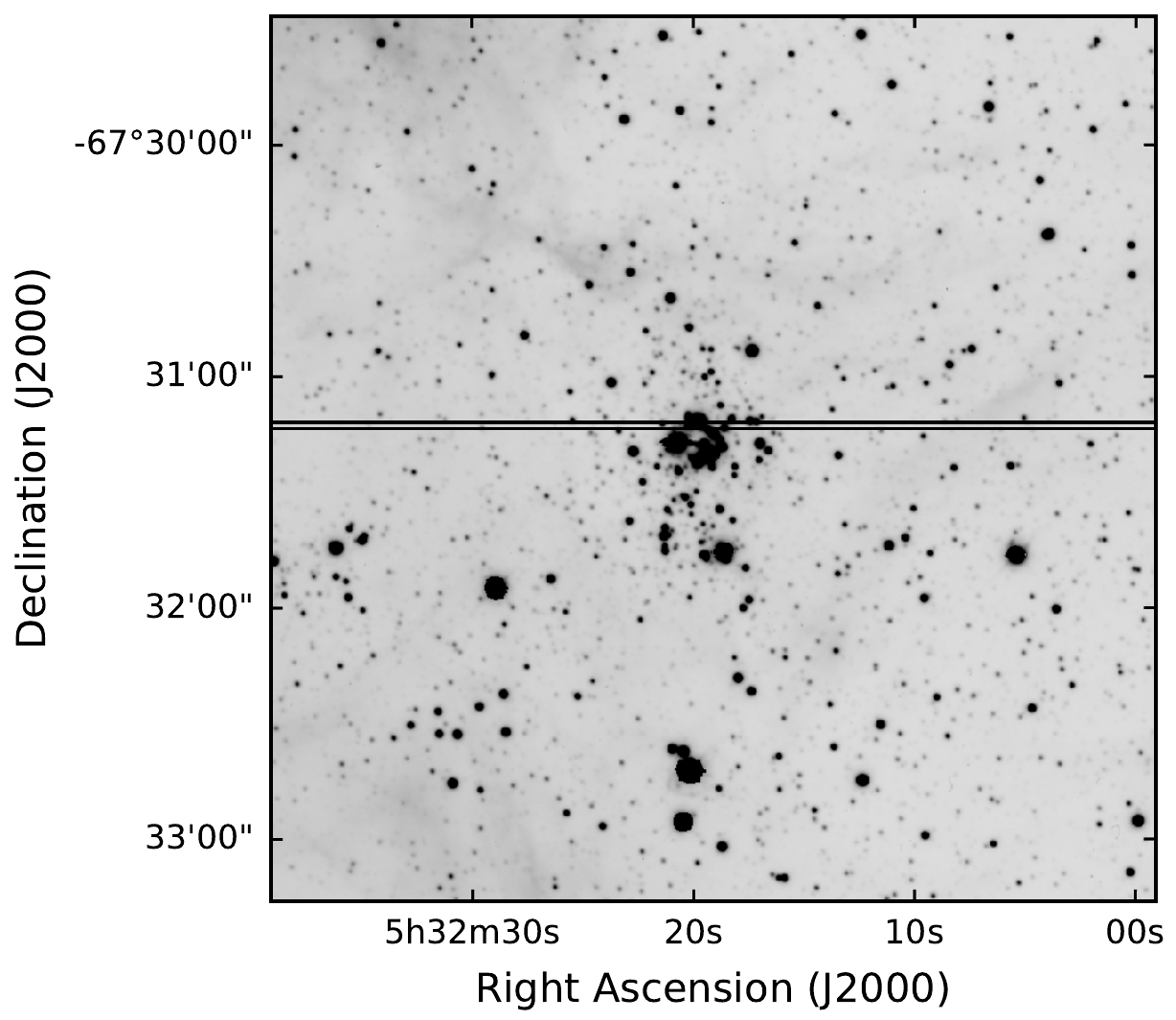}
\label{figure:slit_positions}
}
\hspace{1.2cm}
\subfigure[]{
\hspace{-1.cm}
\vspace{-0.3cm}
\includegraphics[scale=0.37]{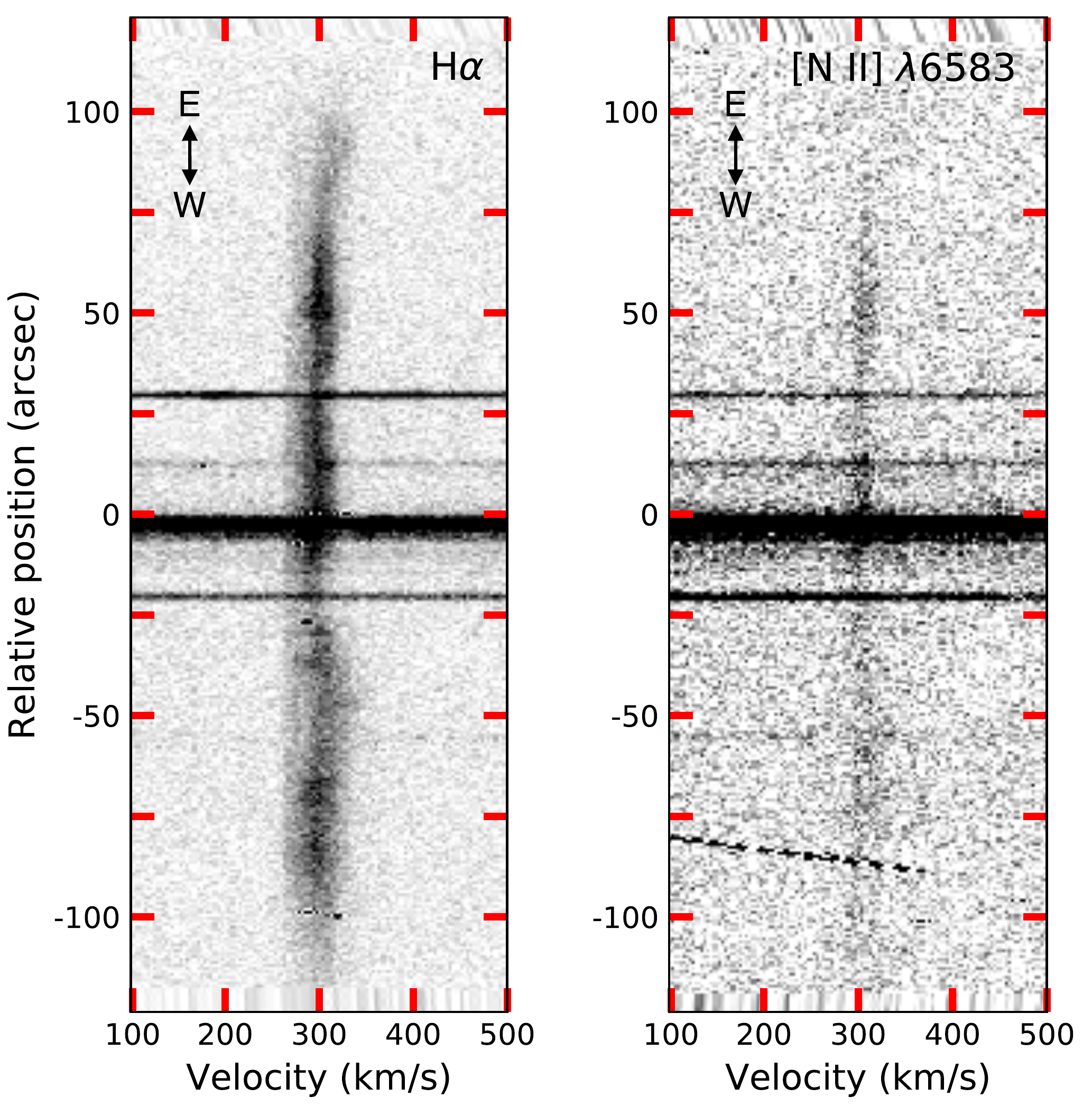}
\label{figure:echelle}
}
\end{tabular}
\caption{(a) The 4\,m MOSAIC H$\alpha$ image of \snr{0532} marked with the CTIO echelle slit position. 
(b) H$\alpha$ and [\ion{N}{2}] echellograms of the slit position marked in (a). The slit orientation is 
marked in the upper left corner.  Note that the lack of signal at the top and bottom 
ends of the slit is caused by vignetting in the optics of the echelle spectrograph.
}
\label{figure:slit}
\end{figure*}

\subsection{Multiwavelength Morphology of the SNR}

SNR B0532$-$67.5 was first identified by its shell-like nonthermal radio emission
\citep{Mathewson1985}.  
The ASKAP 888 MHz continuum image in Figure \ref{figure:snr0532_line_images} 
indeed exhibits a complete shell structure with the eastern side significantly 
brighter than the western side.  
Superposed on the south rim of the radio shell is a bright compact source that 
corresponds to the cataloged radio source SUMSS J053215$-$673308 \citep{mauch2003}.
This compact radio source has no X-ray counterpart; furthermore, three other 
comparable compact radio sources without obvious counterparts are observed within 
10$'$ from the SNR.
It is likely that the compact radio source on the south rim of SNR B0532$-$67.5 and
the other three compact sources are all associated with background quasars or 
radio galaxies. This compact radio source will not be discussed further in this paper.

XMM-Newton X-ray image of SNR B0532$-$67.5 exhibits a shell qualitatively similar 
to the radio shell but different in surface brightness distribution.  
For easier comparisons between radio and X-ray emissions, we have plotted 
X-ray contours over the radio image in Figure \ref{figure:snr0532_line_images}.
The X-ray surface brightness of SNR B0532$-$67.5 is the highest in the northeast 
quadrant and not detected in the southwest quadrant.
The outer boundary of the X-ray emission is similar to that of the radio emission,
except where the X-ray emission is not detected.
In the northwest quadrant, the X-ray image shows a simple rim, while the radio
image shows a bright ridge that contributes to the ring morphology, and exterior 
to the ridge is a fainter extension that is coincident with the X-ray rim. 

SNR B0532$-$67.5 is well known for its lack of an optical shell structure.
As shown in Figure \ref{figure:snr0532_line_images}, no shell structure
is seen in either H$\alpha$ or forbidden line images.
However, when we plot the radio contours over the H$\alpha$ image, we can see 
enhanced H$\alpha$ emission along the brightest portion of the radio shell rim,
especially in the north and southeast directions.
These H$\alpha$ emission regions may be associated with the SNR, but may not 
be compressed by strong shocks, since no narrow filaments are seen.

The \ion{H}{1} map provides the best depiction of the large-scale interstellar
environment of SNR B0532$-$67.5.  
Figure~\ref{figure:hispec}a shows an \ion{H}{1} column density map extracted from
the ATCA+Parkes data cube \citep{Kim2003} overplotted with the ASKAP 888 MHz 
continuum contours. 
It is quite evident that the SNR is in a cavity that is connected with the 
much larger cavity in the supergiant shell LMC-4 to the north \citep{Meaburn1980, Book2008}.
The position-velocity plot in Figure~\ref{figure:hispec}b shows even more
unambiguously that the cavity is a true void.
A close-up comparison between the ASKAP 888 MHz continuum image and the ATCA+Parkes
\ion{H}{1} column density map is presented in Figure~\ref{figure:hiaskap}.
It is clear that the brightest radio and X-ray emission originates from areas where 
the SNR meets the \ion{H}{1} cavity wall.

\subsection{Hot Gas in the SNR Interior}

The XMM-Newton X-ray observation of SNR B0532$-$67.5 has been 
analyzed by \citet{Maggi2016}.  They used non-equilibrium ionization
(NEI) models to fit the integrated X-ray spectrum of the SNR, 
and the best-fit model has a plasma temperature of 
$kT= 0.53^{+0.29}_{-0.11}$ keV and an emission measure of 
2.3$^{+3.53}_{-1.34}$ $\times10^{58}$ cm$^{-3}$.
The emission measure is $\int N_{\rm e}^2 dV$, where $N_{\rm e}$ is the 
electron density and $V$ is the emitting volume.
Assuming a uniform hot gas with a filling factor of $f$, the emission
measure can be expressed as $N_{\rm e}^2 V f$.
Adopting their plasma parameters and X-ray size of the SNR (285$''$ diameter), 
and a a helium to hydrogen number density ratio $N_{\rm He}/N_{\rm H}$ = 0.1,
we derive an rms electron density of 0.064$^{+0.037}_{-0.023} f^{-1/2}$ cm$^{-3}$.
The total hot gas mass is $1.17 N_{\rm e} m_{\rm H} V f$, where $m_{\rm H}$ is
the mass of a hydrogen atom, and is 350$^{+200}_{-125} f^{1/2} M_\odot$ for 
SNR B0532$-$67.5.
This large mass indicates that the hot gas is dominated by the swept-up 
interstellar medium (ISM).
The total thermal energy in the hot gas is (3/2)$N k T V f$, where 
$N \sim N_{\rm e} + N_{\rm H} + N_{\rm He} \sim 1.92N_{\rm e}$,
and is $4.5^{+3.2}_{-1.9} \times 10^{50} f^{1/2}$ ergs for SNR B0532$-$67.5.

SNR B0532$-$67.5 is in a low-density environment.  It would be of interest
to compare its hot gas properties to those of SNRs in a normal diffuse 
medium, such as the N206 SNR, which has also been analyzed by \citet{Maggi2016}.
The plasma temperature of N206, $kT = 0.52^{+0.05}_{-0.04}$ keV, is very similar 
to that of B0532$-$67.5; with a size of 180$''$, N206 is smaller than B0532$-$67.5,
and the emission measure of N206, 1.63$^{+0.21}_{-0.16}$ $\times10^{58}$ cm$^{-3}$,
is also lower than B0532$-$67.5.  Using these plasma parameters of N206,
we find that N206 SNR's hot gas density is $\sim$1.7 times as high as that of SNR 
B0532$-$67.5, but the total thermal energy is only 0.43 times that of 
B0532$-$67.5. Apparently the presence or absence of an H$\alpha$ shell of a SNR
does not cause more than a factor of $\sim$2 effects on the hot gas properties.

\subsection{Kinematics of the Superposed Interstellar Gas}

We use the long-slit echelle observation to probe the warm (10$^4$ K) ionized gas
superposed on SNR B0532$-$67.5.
Figure \ref{figure:slit} shows the slit position marked on the MOSAIC H$\alpha$
image and echelle images of the H$\alpha$ and [\ion{N}{2}] $\lambda$6583 lines.
Although the H$\alpha$ image does not show much appreciable nebular emission, the
echelle image detects the H$\alpha$ emission nicely.  
The H$\alpha$ line shows a narrow component at a constant velocity of $\sim$275 km s$^{-1}$
and a broad irregular component.  The narrow component is actually the telluric OH 6-1 P2(3.5)
line at 6568.779 \AA\ \citep{Osterbrock1996}.
Only the broad component originates from ionized gas in the LMC and is relevant here.
This broad component has a great deal of velocity and width variations along the slit.
Overall the H$\alpha$ line has a centroid near 300 km s$^{-1}$, similar
to the bulk \ion{H}{1} gas associated with the N57 \ion{H}{2} complex to the south of 
the SNR.  The broadest wings of the H$\alpha$ line has a full-width-at-zero-intensity 
approaching 80 km s$^{-1}$.  Such line profiles are 
quite common for diffuse ionized ISM in the disk of a galaxy, and the line broadening
is caused by a combination of large-scale motion and turbulence. An example of such 
broad line profile can be seen in the slit position EII-5 outside the supergiant
shell LMC-2 reported by \citet{Points1999}.

\begin{figure}[h]  
\begin{center}
\epsscale{1.2} 
\plotone{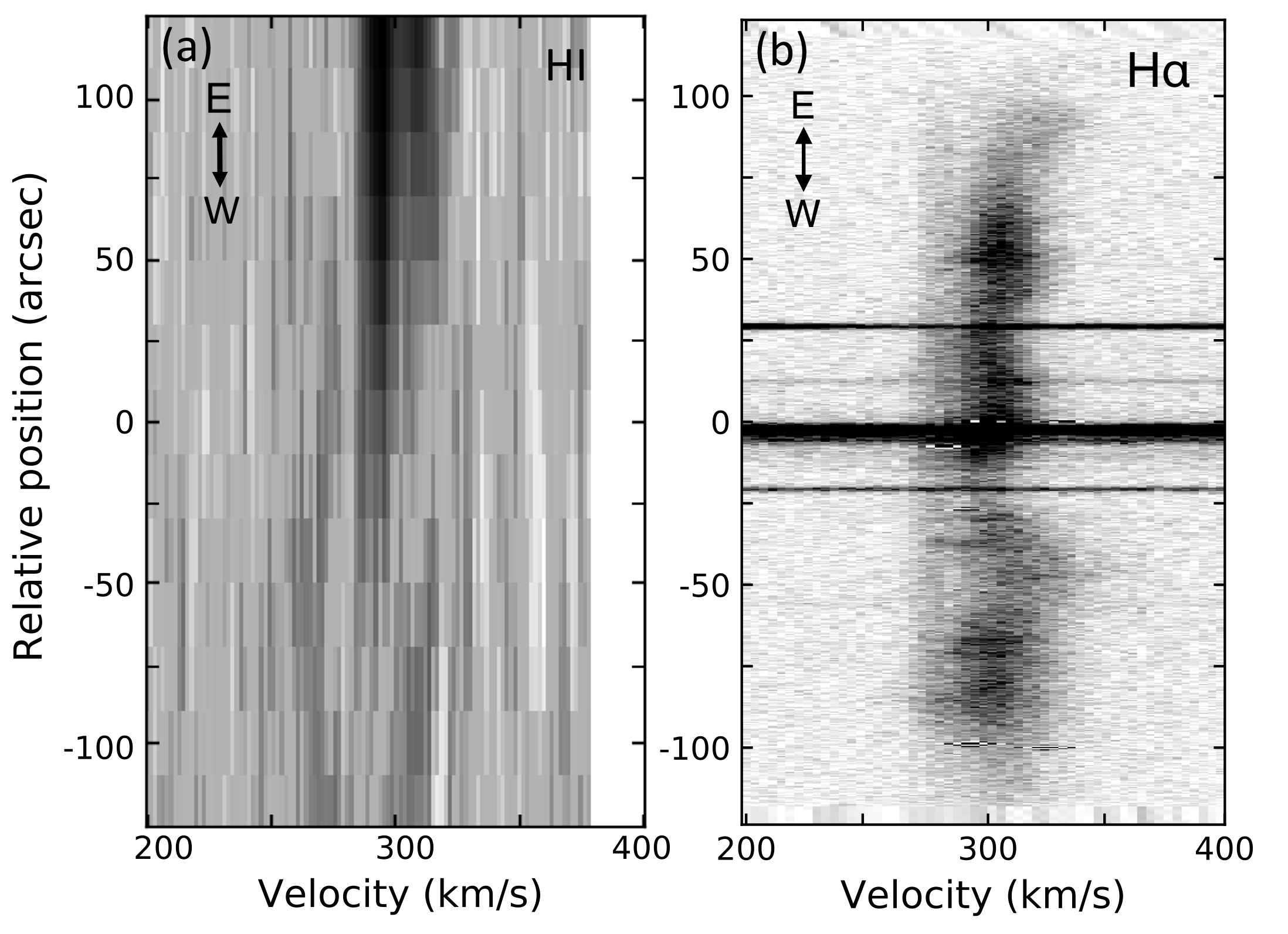} 
\caption{(a) \ion{H}{1} position-velocity diagram extracted along the echelle slit position
shown in Figure~\ref{figure:slit}a.
(b) Long-slit echelle image of the H$\alpha$ line rescaled for direct comparison with the \ion{H}{1}
velocity-position diagram in panel (a). Note that the narrow component at a constant velocity of
$\sim$275 km s$^{-1}$ is the telluric OH 6-1 P2(3.5) line.  Only the broad component is valid 
emission from ionized gas in the LMC.}
\label{figure:hiha}
\end{center}
\end{figure}

We have extracted an \ion{H}{1} position-velocity diagram along the echelle slit
position and present it along the H$\alpha$ echellogram in Figure~\ref{figure:hiha}.
The velocity structures in the \ion{H}{1} gas and \ion{H}{2} gas are quite different.
The \ion{H}{1} 21 cm line shows line splittings that are most likely associated with
large-scale structures, while the H$\alpha$ line shows broad asymmetric velocity profiles
without line splitting. Considering that the isothermal sound velocity of 10$^4$ K ionized
gas is $\sim$10 km~s$^{-1}$ and that the width of the H$\alpha$ line is quite supersonic,
the absence of line splitting indicates that the motion is dominated by turbulence rather
than organized expansion. Evidently the ionized \ion{H}{2} gas and the neutral 
\ion{H}{1} gas do not have the same kinematic origin; furthermore, the SNR most likely
has impacted the \ion{H}{2}, rather than the \ion{H}{1}, kinematics.
It is not clear whether and how the velocity structure of the H$\alpha$ line
is affected by the SNR B0532$-$67.5.


\subsection{Magnetic Field in the SNR} 

\begin{figure*}[th]  
\begin{center}
\epsscale{1.2} 
\hspace{-0.5cm}
\plotone{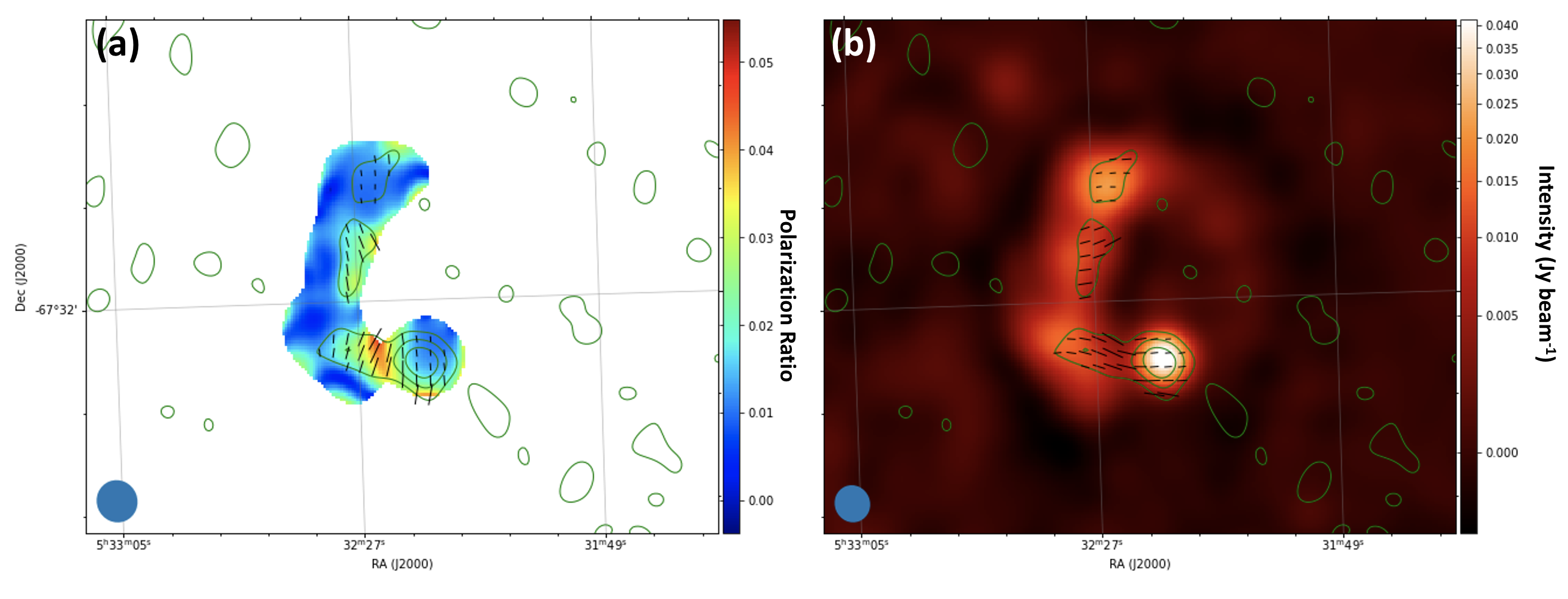} 
\caption{(a) ASKAP 888 MHz continuum degree of polarization (color), polarized emission (green contour), and polarized vector (black vector). The contour levels are 1.5, 2.5, 3.5$\sigma$ where $1\sigma$ corresponds to 0.14 mJy beam$^{-1}$. The vector length denotes degree of polarization and the blue filled ellipse at the bottom left corner denotes the beam size, $35\arcsec\times35\arcsec$. (b) ASKAP 888 MHz continuum stoke I emission (color), polarized emission (green contour), and polarized vector rotated by 90$\arcdeg$ (black vector). The contour levels, vector length, and the beam size are the same as those in panel (a). }
\label{figure:askappol}
\end{center}
\end{figure*}

The ASKAP 888 MHz images in Stokes I, Q, U, and V can be used to probe the magnetic field
in B0532$-$67.5. We have smoothed these images to a beam size of $35\arcsec\times35\arcsec$, 
averaged images in respective Stokes parameters, and made an image cube of these four 
averaged maps. 
Figure \ref{figure:askappol}a shows an image of the degree of polarization ($\sqrt{Q^2+U^2} / I$)
superposed by contours of polarized emission ($\sqrt{Q^2+U^2}$) and polarization vectors.  
Figure \ref{figure:askappol}b shows the Stokes I image (total intensity) 
superposed by contours of polarized emission. 
As the magnetic field direction is perpendicular to the polarization vector, we rotate 
the polarization vector by 90$\arcdeg$ and plot them over the image in 
Figure~\ref{figure:askappol}b. 

It is interesting to note that the polarized emission is stronger toward the inner wall of 
the radio shell, where the \ion{H}{1} column density is lower (see Fig.~\ref{figure:hiaskap}).
The large size of SNR B0532$-$67.5 implies that it is old.
In evolved old SNRs, the SNR shell consists of the ambient ISM compressed by radiative 
shocks \citep{dubner2015}, and thus the magnetic field is expected to be tangential to 
the SNR shell \citep[e.g.][]{milne1968, milne1987, landecker1999, wood2008, dubner2015}.
In SNR B0532$-$67.5 polarized radio emission is detected only in the bright eastern half
of the shell. The magnetic field direction derived from the polarization vectors is parallel
to the shell rim in the northern and southern parts of the polarized component, but appears to
be perpendicular to the shell in the middle.
This mixed magnetic field directions indicate that the polarization vectors could have been
affected by foreground Faraday rotation.  Radio observations of SNR B0532$-$67.5 in more 
wavelengths are needed to investigate the magnetic field and foreground Faraday rotation.


\section{Summary and Conclusions} \label{sec:discussion}  

The LMC is an ideal galaxy where SNRs can be studied in great detail; furthermore,
the resolved underlying stellar population can be used to infer properties of SN 
progenitors, especially for the core-collapse SNe.
This paper reports a detailed study of the SNR B0532$-$67.5.  Its SNR nature
was diagnosed by its nonthermal radio emission and confirmed by its diffuse 
X-ray emission; however, it does not exhibit any optical shell structure in 
either H$\alpha$ or forbidden lines.

We have analyzed the $V$ versus $B-V$ CMD of stars brighter than $V = 17$ mag in and
around SNR B0532$-$67.5, compared their locations in the CMD with evolutionary 
tracks to assess their initial masses, compared them with isochrones to estimate
their ages, and examined their spatial distribution relative to the core of LH75.
We find a segregation in stellar masses -- the most massive stars (20-30 $M_\odot$) 
are distributed to the southwest and outside of LH75, while the stars in the core of 
LH75 have masses up to $\sim$15 $M_\odot$.  Comparisons with isochrones indicate that
LH75 is 20--25 Myr old, while the massive stars to its southwest are less than 10 Myr 
old.  It is possible that the energy feedback of LH75 compressed the ambient ISM and
formed the massive stars to its southwest.
The center of SNR B0532$-$67.5 is closer to the core of LH75 than the younger massive
stars; therefore, we suggest that the SN progenitor was a member of LH75 and its initial
mass was $\sim$15 $M_\odot$.
The low ambient interstellar density is responsible for the SNR's lacking a visible
shell structure in H$\alpha$ and forbidden lines.

XMM-Newton X-ray observations SNR B0532$-$67.5 were analyzed by \citet{Maggi2016},
and we use their best-fit model to the X-ray spectrum to estimate the hot gas mass 
and thermal energy in the SNR interior.  Assuming a filling factor of 0.1, the hot gas
mass is $111^{+63}_{-40} M_\odot$ and the total thermal energy in the SNR interior
is $1.4^{+1.0}_{-0.6} \times 10^{50}$ ergs, about 1/7 of a canonical SN explosion 
energy of 10$^{51}$ ergs. This amount of thermal energy is 
within a factor of 2--3 similar to those in SNRs with visible H$\alpha$ shells of
comparable SNR sizes, such as N206 and DEM\,L316A and DEM\,L316B \citep{Williams2005, WC2005}.

The ATCA + Parkes \ion{H}{1} data cube of the LMC \citep{Kim2003} is used to view
the large-scale interstellar environment.  SNR B0532$-$67.5 is associated with LH75 in the
south end of the supergiant shell LMC-4 \citep{Meaburn1980}.  
The SNR is projected in an \ion{H}{1} cavity with the brightest radio and X-ray rim 
abutting the \ion{H}{1} wall. The \ion{H}{1} cavity is apparently connected with the 
central cavity of the supergiant shell LMC-4 \citep{Book2008}.
This large-scale structure of \ion{H}{1} amply demonstrates an intricate relationship between
the ISM structure and star formation history, and whether a core-collapse SN can produce an
SNR with detectable optical signatures critically depends on the density of the ambient medium,
which in turn depends on the star formation history and energy feedback.

\acknowledgments
This research is supported by the MOST grants 109-2112-M-001-040, 109-2811-M-001-545,
110-2112-M-001-020, and 110-2811-M-001-567 from the Ministry of Science and Technology 
of Taiwan.\\

The Australian SKA Pathfinder is part of the Australia Telescope National Facility which is managed by CSIRO. Operation of ASKAP is funded by the Australian Government with support from the National Collaborative Research Infrastructure Strategy. ASKAP uses the resources of the Pawsey Supercomputing Centre. Establishment of ASKAP, the Murchison Radio-astronomy Observatory and the Pawsey Supercomputing Centre are initiatives of the Australian Government, with support from the Government of Western Australia and the Science and Industry Endowment Fund. We acknowledge the Wajarri Yamatji people as the traditional owners of the Observatory site.\\

\emph{Software:} SAOImage DS9 \citep{Joye2003}, astropy \citep{Astropy2018}, matplotlib \citep{Hunter2007}, numpy \citep{vanderWalt2011, Harris2020}, scipy \citep{Virtanen2020}





\clearpage
\setlength{\tabcolsep}{4pt}
\tabletypesize{\tiny}
\begin{deluxetable*}{cccccccccccc}
\tablecolumns{12}
\tablecaption{ Brightest Massive Stars in and near SNR \snr{0532} from the MCPS data }
\tablehead{ Star & R.A. & Decl. & $U$ & $B$ & $V$ & $I$ & $J$ & $H$ & $Ks$ & Spectral & Mass\\
 & (J2000) & (J2000) & & & & & & & & Type & (\msun)}
\startdata
A & 05:31:59.88 & -67:32:54.92 & 12.54 $\pm$ 0.03 & 13.56 $\pm$ 0.03 & 13.74 $\pm$ 0.13 & 14.01 $\pm$ 0.04 & 14.30 $\pm$ 0.01 & 14.35 $\pm$ 0.02 & 14.41 $\pm$ 0.02 & O9.5\,III$^a$ & $\sim$20 \\
B & 05:32:00.24 & -67:33:08.32 & 14.77 $\pm$ 0.06 & 15.44 $\pm$ 0.41 & 15.73 $\pm$ 0.03 & 15.96 $\pm$ 0.03 & 16.21 $\pm$ 0.02 & 16.23 $\pm$ 0.02 & 16.36 $\pm$ 0.08 & B2\,V$^b$ & 10--15 \\
C & 05:32:04.85 & -67:33:33.80 & 12.32 $\pm$ 0.01 & 13.33 $\pm$ 0.01 & 13.56 $\pm$ 0.01 & 13.71 $\pm$ 0.06 & 14.00 $\pm$ 0.01 & 14.06 $\pm$ 0.01 & 14.11 $\pm$ 0.02 & B0\,II$^a$ & 30--35 \\
D & 05:32:05.47 & -67:31:46.88 & 16.29 $\pm$ 0.06 & 14.62 $\pm$ 0.01 & 13.03 $\pm$ 0.01 & - & \, 10.19 $\pm$ 0.02$^c$ & \quad 9.42 $\pm$ 0.03$^c$ & \quad 9.19 $\pm$ 0.02$^c$ & K4\,I$^b$ & 13--15 \\
E & 05:32:09.63 & -67:34:13.08 & 14.47 $\pm$ 0.03 & 15.41 $\pm$ 0.24 & 15.73 $\pm$ 0.07 & 15.76 $\pm$ 0.06 & 16.19 $\pm$ 0.03 & 16.25 $\pm$ 0.03 & 16.39 $\pm$ 0.09 & B0--1\,V$^b$ & 10--15 \\
F  &  05:32:11.51  &  -67:32:30.08  &  13.82  $\pm$  0.08  &  14.79  $\pm$  0.01  &  15.01  $\pm$  0.04  &  15.23  $\pm$  0.03  &  15.45  $\pm$  0.01 & 15.53 $\pm$ 0.02 & 15.55 $\pm$  0.04 & B0\,V$^b$  &  $\sim$20    \\
\,\,\,G$^d$ & 05:32:17.33 & -67:30:53.42 & 13.11 $\pm$ 0.03 & 14.01 $\pm$ 0.02 & 14.32 $\pm$ 0.02 & 13.87 $\pm$ 0.04 & 13.92 $\pm$ 0.01 & 13.85 $\pm$ 0.01 & 13.70 $\pm$ 0.02 & - & - \\
H &  05:32:18.66  &  -67:33:01.70  &  13.53  $\pm$  0.04  &  14.40  $\pm$  0.01  &  14.62  $\pm$  0.03  &  14.76  $\pm$  0.03  & 14.87 $\pm$ 0.01 & 14.82  $\pm$  0.01 & 14.79  $\pm$  0.03 & -  & $\sim$20 \\
I & 05:32:18.67 & -67:31:45.98 & 13.04 $\pm$ 0.01 & 13.49 $\pm$ 0.01 & 12.70 $\pm$ 0.01 & - & \quad 9.71 $\pm$ 0.02$^c$ & \quad 8.94 $\pm$ 0.03$^c$ & \quad 8.64 $\pm$ 0.03$^c$ & - & $\sim$12  \\
\,\,\,J$^e$ & 05:32:19.80 & -67:31:11.46 & 11.82 $\pm$ 0.10 & 12.51 $\pm$ 0.11 & 12.18 $\pm$ 0.07 & 12.53 $\pm$ 0.15 & 15.23 $\pm$ 0.03 & 14.78 $\pm$ 0.11 & - & - & 12--15 \\
 &  &  &  &  &  &  & 14.61 $\pm$ 0.07 & 14.18 $\pm$ 0.04 & 13.92 $\pm$ 0.07 &  &  \\
 &  &  &  &  &  &   & 14.72 $\pm$ 0.06 & - & - &  &  \\
 &  &  &  &  &  &  & 15.41 $\pm$ 0.06 & 15.45 $\pm$ 0.07 & 15.44 $\pm$ 0.13 &  &  \\
K & 05:32:19.82 & -67:31:22.12 & 12.37 $\pm$ 0.07 & 13.28 $\pm$ 0.08 & 13.20 $\pm$ 0.13 & - & 13.87 $\pm$ 0.02 & - & - & - &  $\sim$10 \\
L & 05:32:20.16 & -67:32:42.04 & - & 15.52 $\pm$ 0.02 & 12.98 $\pm$ 0.12 & \quad 9.7$^f$ & \quad 9.22 $\pm$ 0.03$^c$ & \quad 8.37 $\pm$ 0.03$^c$ & \quad 8.01 $\pm$ 0.04$^c$ & M2\,I$^f$ & $\sim$19 \\ 
M & 05:32:20.55 & -67:32:55.50 & 13.94 $\pm$ 0.01 & 13.54 $\pm$ 0.01 & 12.96 $\pm$ 0.01 & - & 11.77 $\pm$ 0.01 & 11.50 $\pm$ 0.01 & 11.45 $\pm$ 0.02 & F8\,I$^b$ & 10--12 \\
N  &  05:32:20.92  &  -67:32:36.38  &  13.51  $\pm$  0.29  &  14.41  $\pm$  0.01  &  14.62  $\pm$  0.03  &  14.94  $\pm$  0.23  & -  &  15.26 $\pm$ 0.06 & -  & -  & $\sim$20 \\
O & 05:32:21.16 & -67:31:34.50 & 15.10 $\pm$ 0.08 & 15.85 $\pm$ 0.06 & 16.13 $\pm$ 0.04 & 15.96 $\pm$ 0.08 & 16.62 $\pm$ 0.03 & 16.66 $\pm$ 0.04 & 16.71 $\pm$ 0.11 & B2\,V$^b$ & 10--12 \\
P & 05:32:26.41 & -67:31:52.32 & 14.24 $\pm$ 0.04 & 14.86 $\pm$ 0.02 & 15.26 $\pm$ 0.19 & 15.44 $\pm$ 0.22 & 15.65 $\pm$ 0.01 & 15.70 $\pm$ 0.02 & 15.80 $\pm$ 0.05 & B1\,V$^b$ & $\sim$13 \\
Q & 05:32:28.87 & -67:31:54.77 & 10.89 $\pm$ 0.14 & 12.48 $\pm$ 0.28 & 11.76 $\pm$ 0.13 & - & 11.84 $\pm$ 0.01 & 11.81 $\pm$ 0.01 & 11.73 $\pm$ 0.02 & B8\,I$^g$ & \, 15--18 \\
R & 05:32:36.13 & -67:31:44.58 & 14.04 $\pm$ 0.03 & 14.14 $\pm$ 0.02 & 13.61 $\pm$ 0.04 & 12.81 $\pm$ 0.13 & 12.63 $\pm$ 0.01 & 12.35 $\pm$ 0.01 & 12.33 $\pm$ 0.02 & - & \, 7--12 
\enddata
\tablenotetext{a}{From \citet{Conti1986,Massey2002}.}
\tablenotetext{b}{This paper; photometrically estimated.}
\tablenotetext{c}{From 2MASS.}
\tablenotetext{d}{Colors are poorly fitted with stellar atmosphere model in the MCPS data.}
\tablenotetext{e}{Multiple stars. The mass estimated from the CMD and stellar evolutionary tracks 
has large uncertainties and represents an upper limit. }
\tablenotetext{f}{From \citet{Westerlund1981}. }
\tablenotetext{g}{From \citet{Rousseau1978}. }
\label{table:photometry}
\end{deluxetable*}
\clearpage

\clearpage

\noindent {\bf Appendix: Notes on individual stars:}\\

The mass estimates of stars are approximate.  
The uncertainty in the mass estimates may be 2--3 $M_\odot$ for stellar masses
below 10 $M_\odot$, but the uncertainty is much larger for high-mass stars and
may be 5 $M_\odot$ or even higher. \\

\noindent {\bf Star A.} This star was cataloged as BI 192 and classified as OB0 by \citet{Brunet1975}. 
The spectral type was later refined to O9.5\,III \citep{Conti1986,Massey2002}.
This star has evolved off the main sequence (MS) and is located close to the evolutionary 
track of a 20 $M_\odot$ star; therefore, we assign an initial mass of $\sim$20 $M_\odot$. \\

\noindent {\bf Star B.} This star has a large uncertainty in its $B$ magnitude. Its colors in other passbands
are similar to a B2\,V star.  Within its $B-V$ uncertainty, its location in the CMD could overlap the evolutionary tracks of 10-15 $M_\odot$ star; thus we assign a stellar mass of 10-15 $M_\odot$. \\

\noindent {\bf Star C.} This star was cataloged as BI 193 and classified as OB0 by \citet{Brunet1975}. 
The spectral type was later refined to B0\,II \citep{Conti1986,Massey2002}. Its
location in the CMD is between the evolutionary tracks of 25 and 40 $M_\odot$ stars.  We assign an initial mass 
of 30-35 $M_\odot$ to this star. \\

\noindent {\bf Star D.} This star was cataloged as W61 19-1 and classified as M star with large 
uncertainty by \citet{Westerlund1961}.
We find that the colors and magnitudes of this star are similar to a K4\,I star, which has an initial mass 
of $\sim$13 $M_\odot$.  This star's location in the CMD is on the evolutionary track of a 15  $M_\odot$ star.  
Thus, we assign an initial mass of 13-15 $M_\odot$ to this star. \\

\noindent {\bf Star E.} The colors and magnitudes of this star suggest a spectral type of B0-1\,V.  Its B magnitude has a larger 
error than the others, which may be the cause of its location bluer than the evolutionary tracks.  Based on its likely 
spectral type of B0-1\,V and its location in the CMD, we assign an initial stellar mass of 10--15 $M_\odot$. \\

\noindent {\bf Star F.} The colors and magnitudes of this star are similar to a B0\,V star.  Its location in the CMD is close to
the evolutionary track of a 20 $M_\odot$ MS star. Thus, we assign an initial stellar mass of $\sim$20 $M_\odot$.  \\

\noindent {\bf Star G.} This star was flagged in MCPS for colors poorly fitted 
by atmospheric models, as noted by \citet{Zaritsky2004}.  The combination of the MCPS $UBVI$ and IRSF $JHK_s$ 
photometric data shows clearly that this object has a near-infrared excess,
indicating the existence of a late-type companion.  Spectroscopic observations 
are needed to reveal the nature of this system. \\

\noindent {\bf Star H.} The colors of this star are similar to B1, but its $M_V$ is 1.5 mag brighter than a B1\,V star.
The star is close to the evolutionary track of a 20 $M_\odot$ star.
We estimate a spectral type of B1\,II and assign an initial stellar mass of 20 $M_\odot$. \\

\noindent {\bf Star I.} This star is close to the core of LH75, and was cataloged as 
W61 19-14 without spectral classification. \citep{Westerlund1961}.
This star's colors and magnitudes are complex and must
contain multiple stars.  Its location in the CMD suggests an evolved $\sim$12 $M_\odot$ star.  \\

\noindent {\bf Star J.} This object is at the core of LH75. It has a single 
entry in the MCPS catalog, but the IRSF catalog lists four sources within 1$''$ 
from the MCPS star position.
Assuming that the optical emission is dominated by one star, its location in 
the CMD is between the evolutionary tracks of 12--15 $M_\odot$ stars.  We assign a 
mass of 12--15 $M_\odot$ for the most massive star within this object. \\

\noindent {\bf Star K.} This star is near the crowded core region of LH75, and its ground-based photometric measurements may have 
larger errors due to confusion.  Its apparent $UBV$ colors do not match any single MS or supergiant stars. 
If the photometric measurements indeed correspond to a single star, its location in the CMD indicates that it is an
evolved star with an initial mass $\sim$10 $M_\odot$. \\

\noindent {\bf Star L.} This star was cataloged as W61 19-24 by \citet{Westerlund1961},
and as WOH S 383 by \citet{Westerlund1981}, who also classified the star to be an M2\,I 
red supergiant.  With a $B-V = 2.54\pm 0.14$, this star is off the 
chart in Figure~\ref{figure:CMD_snr0532_MCPS}.    
The mass of an M2\,I star would be $\sim$19 $M_\odot$. \\

\noindent {\bf Star M.}  The colors of this star are consistent with an F8\,I, but about 1 mag fainter.  Its location
in the CMD is near the evolutionary tracks of 10 and 12 $M_\odot$ stars.  We assign a stellar mass of 10--12 $M_\odot$.\\

\noindent {\bf Star N.} The colors of this star are similar to B1, but its $M_V$ is 1.5 mag brighter than a B1\,V star.
The star is close to the evolutionary track of a 20 $M_\odot$ star.
We estimate a spectral type of B1\,II and assign an initial stellar mass of 20 $M_\odot$. \\

\noindent {\bf Star O.}  The $UBV$ colors suggest a B2\,V spectral type, but about 1 mag brighter.  The uncertainties in colors cannot exclude an earlier spectral type of B1\,V. The star
is probably 10--12 $M_\odot$. \\

\noindent {\bf Star P.}  The $UBV$ photometry of this star has a large uncertainty
in $V$. This large uncertainty is also reflected in its unreasonable location in the CMD. 
These uncertainties may be caused by its being an eclipsing binary \citep{Muraveva2014}.
If we assume the star is on the MS, its $V$ magnitude implies a spectral type 
close to B1\,V, and the stellar mass would be $\sim$13 $M_\odot$. \\

\noindent {\bf Star Q.}  This star, aka HD 269726 and Sk$-$67 178, is visually the 
brightest star within SNR B0532$-$67.5.  
Its colors and magnitudes do not match any standard stars.  It was classified by 
\citet{Rousseau1978} to be of spectral type B8\,I, which would suggest an
initial stellar mass of $\sim$18 $M_\odot$, although the star is closest to
the evolutionary track of a 15 $M_\odot$ star in Figure \ref{figure:CMD_snr0532_MCPS}. \\

\noindent {\bf Star R.} The colors and magnitudes of this star do not match any main 
sequence or supergiant stars.  Its location in the CMD, 
compared with the stellar evolutionary tracks, suggests that it is an evolved star 
with an initial mass of 7--12 $M_\odot$.

\end{document}